\documentclass[11pt]{article}

\usepackage{amsmath,amssymb,amsfonts}
\usepackage{graphicx}
\usepackage{booktabs}
\usepackage{bm}
\usepackage{hyperref}
\usepackage{geometry}
\usepackage{graphicx}
\usepackage{subcaption}
\usepackage{booktabs}
 \usepackage{multirow}
 \usepackage{authblk}
\usepackage{cite}
\usepackage{orcidlink}

\usepackage{authblk}
\usepackage{orcidlink}

\title{%
Brouwer Degree of Thermodynamic Multicritical\\
Points in Black Holes}

\author[1]{%
Bidyut Hazarika\,\orcidlink{0009-0007-8817-1945}%
\thanks{E-mail: bidyuthazarika1729@gmail.com}}

\author[1,2]{%
Prabwal Jyoti Phukon\,\orcidlink{0000-0002-4465-7974}%
\thanks{E-mail: prabwal@dibru.ac.in}}

\affil[1]{Department of Physics, Dibrugarh University, Dibrugarh 786004, India}

\affil[2]{Theoretical Physics Division, Centre for Atmospheric Studies, Dibrugarh University, Dibrugarh 786004, India}

\date{}

\date{}

\begin{document}

\maketitle


\begin{abstract}

In this manuscript, we propose a novel topological framework based on the heat capacity of black holes to investigate the topology of thermodynamic multicritical points. We construct a two-dimensional thermodynamic vector field whose isolated zeros correspond to the critical points of the system. The local topology of each critical point is characterized by its Brouwer degree, which serves as a locally conserved topological quantity. Applying this formalism to several AdS black hole solutions, we demonstrate that each system possesses a globally conserved topological charge. Although the number of thermodynamic critical points changes as the thermodynamic parameters are varied, the total topological charge remains invariant throughout the evolution. We show that these topological transitions are governed by the creation or annihilation of topologically neutral defect pairs carrying opposite Brouwer degrees. Our results provide a unified topological framework for understanding the emergence, evolution, and classification of thermodynamic multicritical points in black hole systems.
\end{abstract}


\section{Introduction}

Since the formulation of black hole thermodynamics \cite{Bekenstein:1973ur,Hawking:1974rv,Hawking:1975vcx,Bardeen:1973gs}, the study of black hole phase transitions has attracted considerable attention. The earliest thermodynamic phase transitions identified in black holes include the Davies transition and the Hawking--Page transition \cite{Davies:1989ey,Hawking:1982dh}. Over the past two decades, this field has developed significantly, particularly after the realization that black holes in anti-de Sitter (AdS) spacetime exhibit thermodynamic behavior remarkably similar to that of ordinary thermodynamic systems.

A major breakthrough came with the discovery of Van der Waals-like phase transitions in AdS black holes, where the cosmological constant is interpreted as thermodynamic pressure and its conjugate quantity as thermodynamic volume \cite{Kubiznak:2012wp,Dolan:2010ha}. This viewpoint gave rise to the subject of black hole chemistry \cite{Kubiznak:2016qmn}. More recently, black hole phase transitions have also been investigated within the frameworks of restricted phase space thermodynamics and holographic thermodynamics \cite{rp1,rp2,rp3,rp4,rp5,rp6,rp7,rp8,rp9,rp10,rp11} which results into exotic phenomena, such as triple points, reentrant phase transitions, and superfluid-like critical behavior, being reported in various AdS black hole systems \cite{Altamirano:2013uqa,Hennigar:2017apu}.

A recent development in black hole thermodynamics is the discovery of thermodynamic multicriticality, where more than three phases coexist at a single critical point. Although multicritical behavior is well known in condensed matter physics, it has only recently been observed in a few AdS black hole solutions, particularly in higher-dimensional gravity and nonlinear electrodynamics. Only a limited number of four-dimensional black hole solutions are currently known to exhibit such behavior. These discoveries have motivated further investigations into the classification and underlying structure of thermodynamic multicritical points in black holes.\\

Our work is motivated by the recently developed topological approach to black hole thermodynamics, in which a two-dimensional vector field is constructed such that its  defect points correspond to physically significant quantities, such as black hole equilibrium states or thermodynamic phase transition points.  Using this method, the stability and topological classification of black hole solutions have been investigated in Refs.~\cite{t1,t2,t3}. The topology of Hawking--Page and Davies-type phase transitions has also been studied within this framework \cite{t4,t5}. More recently,  these methods have been applied to investigate the topology of critical points in holographically dual boundary matrix models \cite{t6}. Beyond black hole thermodynamics, similar topological techniques have been employed to study various phenomena in black hole spacetimes. In particular, they were first applied to investigate the topology of light rings and timelike circular orbits around black holes \cite{PRL119-251102, PRL124-181101, PRD102-064039, PRD103-104031, PRD105-024049, PRD108-104041, 2401.05495, PRD107-064006, JCAP0723049, 2406.13270}.\\
\subsection{Motivation}
In this manuscript, our main motivation is to develop a topological framework that assigns a Brouwer degree to thermodynamic multicritical points, allowing them to be distinguished and classified from a topological perspective. While recent studies have successfully classified Hawking--Page, Davies-type, and Van der Waals critical points using topological methods, the topology of multicritical points has remained  unexplored.  Our aim is to construct a framework that not only identifies multicritical points but also enables us to study their evolution as the thermodynamic parameters are varied. In particular, we investigate how topological defects are created and annihilated during the evolution of the system. We show that, although the number of thermodynamic critical points may change, the total topological charge remains conserved throughout the evolution. This conserved global topological charge provides a  way to distinguish different black hole systems exhibiting multicritical behavior.

\section{Proposed Vector Field}

Most of the existing thermodynamic topological approaches are constructed from  the generalized off-shell free energy.   In contrast, the topology associated with heat-capacity,  which directly characterize thermodynamic criticality, has remained largely unexplored. In this work, we develop a new topological framework based on the heat capacity and use it to investigate the emergence, evolution, and classification of thermodynamic multicritical points in black holes.\\

 A genuine critical point corresponds to the endpoint of the phase transition curve, where distinct thermodynamic phases become indistinguishable and the equation of state satisfies the criticality conditions
\begin{equation}
\left(\frac{\partial P}{\partial r}\right)_T=0,
\qquad
\left(\frac{\partial^2P}{\partial r^2}\right)_T=0.
\end{equation}

We introduce the two-dimensional thermodynamic vector field to detect the critical points and differentiate them topologically. 
\begin{equation}
\boxed{
\bm{\phi}(r,P)=
\left(
\frac{1}{C_P},
\frac{\partial}{\partial r}\left(\frac{1}{C_P}\right)
\right).
}
\end{equation}
The simultaneous zeros of the vector field correspond to the thermodynamic critical points of the black-hole system. For black holes exhibiting the standard Van der Waals phase transition, such as the RN AdS black hole, the vector field possesses a single thermodynamic critical point. In contrast, black holes showing more then three phase structures,  generally admit several isolated zeros of the vector field, each representing a distinct thermodynamic critical point. Once the critical points are identified as isolated topological defects of the vector field, their local topological properties can be characterized by the associated winding number, or equivalently, the topological charge.  Consequently, critical points that may appear thermodynamically similar can nevertheless possess different topological characters. As will be shown in the following sections, this topological characterization provides a framework for distinguishing and classifying multiple critical points in black-hole thermodynamics according to the local structure of the vector field.
\subsection{Winding Number and Brouwer Degree}

The topological nature of each thermodynamic critical point can be characterized by the winding of the vector field around its isolated zero. Let
\begin{equation}
\bm{\phi}=(\phi_1,\phi_2),
\end{equation}
and define the corresponding normalized vector field
\begin{equation}
\hat{\bm{\phi}}
=
\frac{\bm{\phi}}{|\bm{\phi}|},
\qquad
|\bm{\phi}|=
\sqrt{\phi_1^2+\phi_2^2}.
\end{equation}

To determine the topological charge, we consider a small closed contour $C$ enclosing a single critical point $(r_c,P_c)$, parameterized as
\begin{equation}
r=r_c+\epsilon\cos\theta,\qquad
P=P_c+\epsilon\sin\theta,
\qquad
0\leq\theta\leq2\pi,
\end{equation}
where the radius $\epsilon$ is sufficiently small such that no other critical point is enclosed. As the contour is traversed once, the normalized vector field rotates continuously, and the total rotation defines the winding number
\begin{equation}
w=\frac{1}{2\pi}
\oint_C
\left(
\hat{\phi}_1\,d\hat{\phi}_2
-
\hat{\phi}_2\,d\hat{\phi}_1
\right).
\end{equation}
Equivalently, if $\Omega(\theta)$ denotes the accumulated deflection angle of the normalized vector field along the contour, then
\begin{equation}
w=
\frac{\Omega(2\pi)-\Omega(0)}{2\pi}.
\end{equation}
The winding number is a global topological invariant that counts the net number of rotations of the vector field around the enclosed defect. A complete counterclockwise rotation corresponds to $w=+1$, whereas a complete clockwise rotation corresponds to $w=-1$.

An equivalent local characterization is provided by the Brouwer degree of the mapping.  To characterize the local structure, we first construct the Jacobian matrix
\begin{equation}
J_{ij}
=
\frac{\partial\phi_i}{\partial x_j},
\qquad
x_j=(r,P),
\end{equation}
or explicitly,
\begin{equation}
J=
\begin{pmatrix}
\displaystyle
\frac{\partial\phi_1}{\partial r}
&
\displaystyle
\frac{\partial\phi_1}{\partial P}
\\[2mm]
\displaystyle
\frac{\partial\phi_2}{\partial r}
&
\displaystyle
\frac{\partial\phi_2}{\partial P}
\end{pmatrix}.
\end{equation}

Expanding the vector field in the neighborhood of an isolated critical point
$(r_c,P_c)$ gives
\begin{equation}
\bm{\phi}
=
J\cdot
\begin{pmatrix}
r-r_c\\
P-P_c
\end{pmatrix}
+
\mathcal{O}\!\left((x-x_c)^2\right),
\end{equation}
showing that the Jacobian completely determines the local topology of the defect.

The determinant of the Jacobian specifies whether the local mapping preserves or reverses orientation. Consequently, the Brouwer degree associated with the critical point is
\begin{equation}
\deg(\bm{\phi},x_c)
=
\mathrm{sgn}\!\left(\det J\right),
\end{equation}
provided the zero is non-degenerate $(\det J\neq0)$. For isolated non-degenerate critical points, the winding number and the Brouwer degree are identical,
\begin{equation}
\boxed{
w
=
\deg(\bm{\phi},x_c)
=
\mathrm{sgn}\!\left(\det J\right).
}
\end{equation}
Therefore, the topological charge of a thermodynamic critical point can be determined either globally from the rotation of the normalized vector field around a closed contour or locally from the sign of the Jacobian determinant. In the present work, we compute the winding number numerically using the deflection-angle method and verify that it agrees exactly with the Brouwer degree obtained from the Jacobian matrix for all the critical points considered.\\

Additional information is contained in the eigenvalues
$\lambda_1$ and $\lambda_2$ of the Jacobian matrix. These eigenvalues determine the local geometry of the vector field in the vicinity of the critical point. If the eigenvalues possess opposite signs,
\begin{equation}
\lambda_1\lambda_2<0,
\end{equation}
the critical point is locally saddle-like, indicating that the vector field contracts along one direction while expanding along the other. In this case,
\begin{equation}
\det J<0,
\end{equation}
and the corresponding Brouwer degree  is
\begin{equation}
w=-1.
\end{equation}

If both eigenvalues are real and have the same sign,
\begin{equation}
\lambda_1\lambda_2>0,
\end{equation}
the critical point is node-like. In particular, if both eigenvalues are negative, the vector field is locally attracted towards the critical point, corresponding to a stable node, whereas positive eigenvalues represent an unstable node. Since
\begin{equation}
\det J>0,
\end{equation}
the associated Brouwer degree  is
\begin{equation}
w=+1.
\end{equation}

A third possibility arises when the eigenvalues form a complex conjugate pair,
\begin{equation}
\lambda_{1,2}=\alpha\pm i\beta,
\end{equation}
where $\alpha,\beta\in\mathbb{R}$. In this case, the thermodynamic flow exhibits a spiral structure around the critical point. If $\alpha<0$, the trajectories spiral towards the critical point, corresponding to a stable spiral, whereas $\alpha>0$ describes an unstable spiral. Since the determinant remains positive,
\begin{equation}
\det J=\lambda_1\lambda_2=\alpha^2+\beta^2>0,
\end{equation}
the Brouwer degree of a spiral defect is also
\begin{equation}
w=+1.
\end{equation}

The global topological property of the thermodynamic system is characterized by the total topological charge, defined as the sum of the Brouwer degrees of all isolated critical points,
\begin{equation}
\boxed{
W_{\rm tot}=\sum_i w_i}
\end{equation}
As will be demonstrated in the following sections, although the number of thermodynamic critical points may change as the thermodynamic parameters are varied, the total topological charge remains conserved for a given black hole system.

\section{Reissner--Nordstr\"om--AdS Black Hole}

Before investigating black holes exhibiting multicritical behavior, we first apply the proposed  framework to the Reissner--Nordstr\"om--AdS (RN-AdS) black hole. The RN-AdS spacetime possesses the standard Van der Waals phase transition structure and admits a single thermodynamic critical point. 
For the RN-AdS black hole, the thermodynamic vector field introduced in the previous section takes the form
\begin{equation}
\phi_1=
\frac{8 \pi  p r^4+3 q^2-r^2}{2 \pi  r^2 \left(8 \pi  p r^4-q^2+r^2\right)},
\end{equation}
and
\begin{equation}
\phi_2=
\frac{-64 \pi ^2 p^2 r^8-2 q^2 r^2 \left(40 \pi  p r^2+3\right)+16 \pi  p r^6+3 q^4+r^4}{\pi  r^3 \left(8 \pi  p r^4-q^2+r^2\right)^2}.
\end{equation}

The simultaneous solution of
\begin{equation}
\phi_1=0,
\qquad
\phi_2=0,
\end{equation}
yields a unique isolated zero of the vector field, corresponding to the thermodynamic critical point of the RN-AdS black hole. The normalized vector field rotates once around this isolated defect, giving a winding number
\begin{equation}
w=-1.
\end{equation}
\begin{figure}[h!]
\centering
\includegraphics[width=0.48\textwidth]{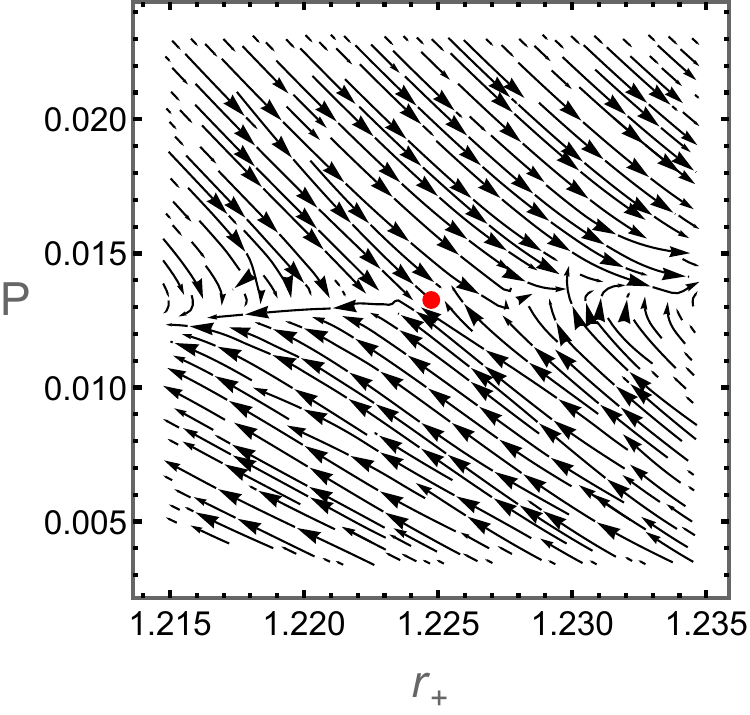}
 \hfill
\includegraphics[width=0.475\textwidth]{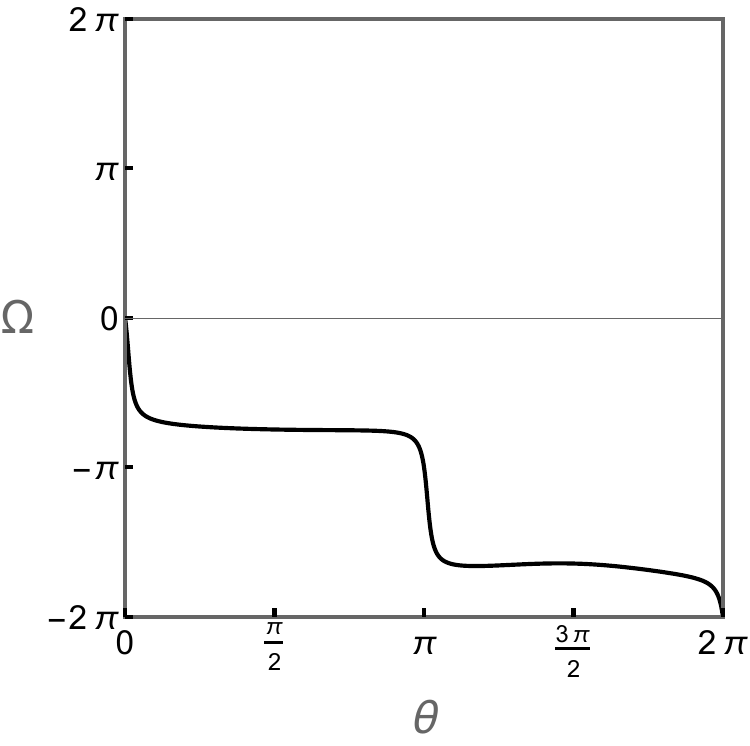}
\caption{Vector plot and inding number for $q=0.5$}
\label{fig:RNTopology}
\end{figure}

The left panel of Figure~\ref{fig:RNTopology} displays the vector field $\phi$ and the red dot represents
the critical point. The flow clearly converges towards a single isolated
topological defect.  Next,  we compute the winding number from
the accumulated deflection angle of the normalized vector field along a small
closed contour surrounding the defect. As shown in the right panel of
Fig.~\ref{fig:RNTopology}(b), the deflection angle decreases continuously by
$2\pi$ after one complete traversal of the contour, corresponding to
\begin{equation}
w=
\frac{\Delta\Omega}{2\pi}
=-1.
\end{equation}

The same result is obtained from the local Jacobian analysis. Evaluating the Jacobian matrix at the critical point, we find
\begin{equation}
\det J<0,
\end{equation}
while its two eigenvalues have opposite signs,
\begin{equation}
\lambda_1<0,
\qquad
\lambda_2>0.
\end{equation}
Thus, the critical point is identified as a saddle-type topological defect with Brouwer degree
\begin{equation}
\deg(\bm{\phi})=-1,
\end{equation}
in complete agreement with the winding number.

This example demonstrates that the proposed vector-field construction correctly reproduces the unique critical point of the RN-AdS black hole and assigns it a well-defined topological charge. 
Although the RN-AdS black hole contains only a single critical point, this
example establishes the methodology employed throughout this work. In the
following section, we apply the same framework to black holes exhibiting
multiple thermodynamic critical points, where the topological characterization
provides a natural means of distinguishing the critical point topologically. 
\section{Euler--Heisenberg Black Hole}

Having established the framework for the
Reissner--Nordstr\"om--AdS black hole, we now extend our analysis to the
Euler--Heisenberg (EH) black hole.  The Euler--Heisenberg black
hole is described by the metric
\begin{equation}
ds^2
=
-f(r)dt^2
+\frac{dr^2}{f(r)}
+r^2(d\theta^2+\sin^2\theta\,d\phi^2),
\end{equation}
with
\begin{equation}
f(r)
=
1-\frac{2M}{r}
+\frac{Q^2}{r^2}
-\frac{aQ^4}{20r^6}
+\frac{8\pi Pr^2}{3},
\end{equation}
where $M$, $Q$, and $a$ denote the black-hole mass, magnetic charge, and
Euler--Heisenberg coupling parameter, respectively.  Here we have substituted $\Lambda=-8\pi P$.
Following the construction introduced in the previous section, the
thermodynamic vector field is defined as
\begin{equation}
\bm{\phi}
=
\left(
\frac{1}{C_P},
\frac{\partial}{\partial r}
\left(\frac{1}{C_P}\right)
\right).
\end{equation}
For the Euler--Heisenberg black hole, the two components of the vector field
become
\begin{equation}
\phi_1=
\frac{
-7aQ^4+12Q^2r^4-4r^6+32\pi Pr^8
}
{
2\pi r^2
\left(
aQ^4+4(-Q^2r^4+r^6+8\pi Pr^8)
\right)
},
\end{equation}
and
\begin{tiny}
\begin{equation}
\phi_2=
\frac{
7a^2Q^8
-8aQ^4r^4
(9Q^2-13r^2-152\pi Pr^4)
-16r^8
\left[
-3Q^4
+Q^2(6r^2+80\pi Pr^4)
+r^4(-1-16\pi Pr^2+64\pi^2P^2r^4)
\right]
}
{
\pi r^3
\left(
aQ^4+4(-Q^2r^4+r^6+8\pi Pr^8)
\right)^2
}.
\end{equation}
\end{tiny}
Unlike the RN-AdS case, depending on the values of the nonlinear coupling $a$ and the
electric charge $Q$,  the thermodynamic phase space admits
two distinct topological defects or critical points.  \\

In the following analysis we fix the nonlinear coupling parameter at
$a=0.04$ and investigate the evolution of the critical points as the charge
$Q$ is varied. The Jacobian matrix, its eigenvalue spectrum, the Brouwer
degree, and the winding number are then employed to classify each isolated
critical point according to its topological character.

\begin{figure}[t]
\centering
\includegraphics[width=0.48\textwidth]{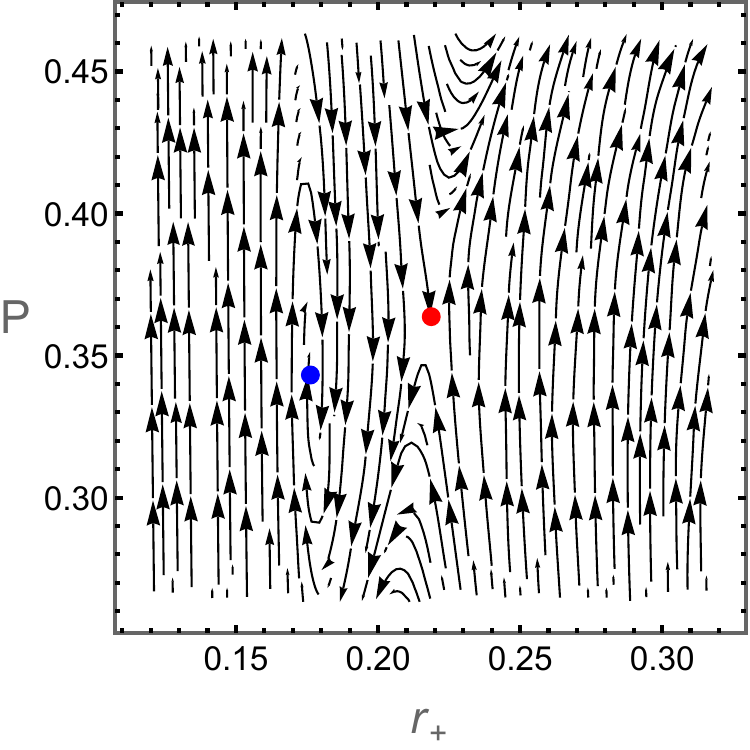}
\hfill
\includegraphics[width=0.48\textwidth]{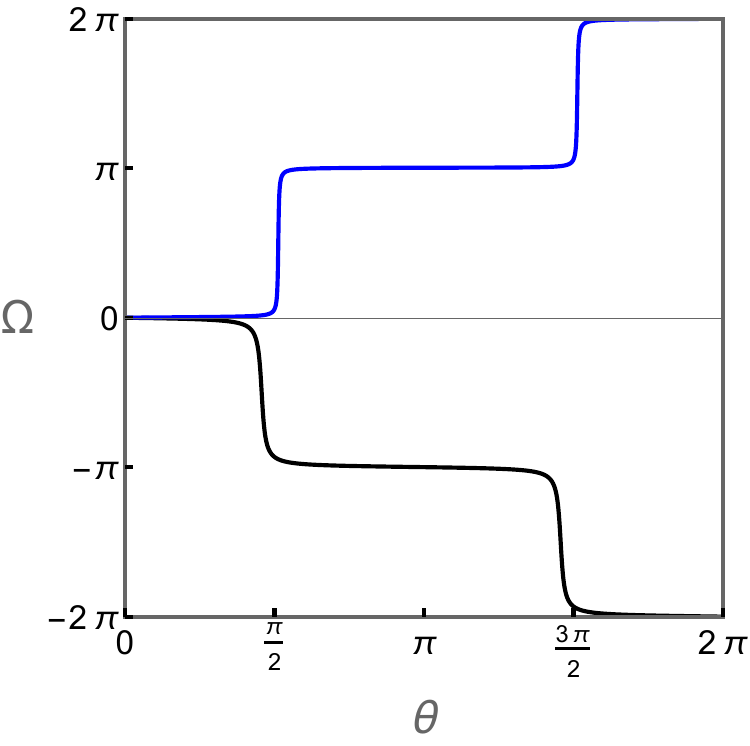}
\caption{
Left: Formation of saddle (red point ) and spiral focus (blue point) topological defect for EH black hole at $a=0.04$ and $Q=0.1$.  Right: Deflection-angle calculation of the winding number
around each critical point. The blue defect has a winding number $w=+1$, whereas the red defect
corresponds to $w=-1$. 
}
\label{fig:EH_vector}
\end{figure}

To illustrate the multicritical structure of the Euler--Heisenberg black hole,
we fix the nonlinear coupling parameter at $a=0.04$ and choose the charge
parameter $Q=0.1$. For these parameter values, the thermodynamic vector field
possesses two isolated zero points,or two distinct
thermodynamic critical points. The corresponding vector field is shown in
Fig.~\ref{fig:EH_vector}.

The winding number associated with each critical point is determined by
evaluating the accumulated deflection angle of the normalized vector field
around a small closed contour enclosing the defect. The blue critical point
is found to possess a winding number
\begin{equation}
w=+1,
\end{equation}
whereas the red critical point carries
\begin{equation}
w=-1.
\end{equation}
The same topological charges are independently reproduced from the Brouwer
degree obtained from the Jacobian matrix. 
The critical point carrying the winding number
\begin{equation}
w=-1
\end{equation}
has a negative Jacobian determinant, and its eigenvalues are found to have
opposite signs,
\begin{equation}
\lambda_1>0,\qquad
\lambda_2<0.
\end{equation}
This indicates that the vector field is locally expanding along one direction
while contracting along the other, identifying the critical point as a
saddle-type topological defect.

In contrast,  The second critical point possesses a positive Jacobian determinant and a
winding number $w=+1$. The eigenvalues of the Jacobian form a complex
conjugate pair,
\begin{equation}
\lambda_{1,2}
=
-11.5536
\pm
46.7831\,i,
\end{equation}
whose negative real part indicates that the vector field spirals towards the
critical point. Consequently, this defect is identified as a stable spiral
(focus) as it is visible in the Fig.~\ref{fig:EH_vector}.  The positive determinant implies that the local
mapping preserves orientation, consistent with its positive Brouwer degree
and winding number.
\subsection{Evolution of the Topological Defects}

\begin{figure*}[t]
\centering

\begin{subfigure}[b]{0.415\textwidth}
    \centering
    \includegraphics[width=\textwidth]{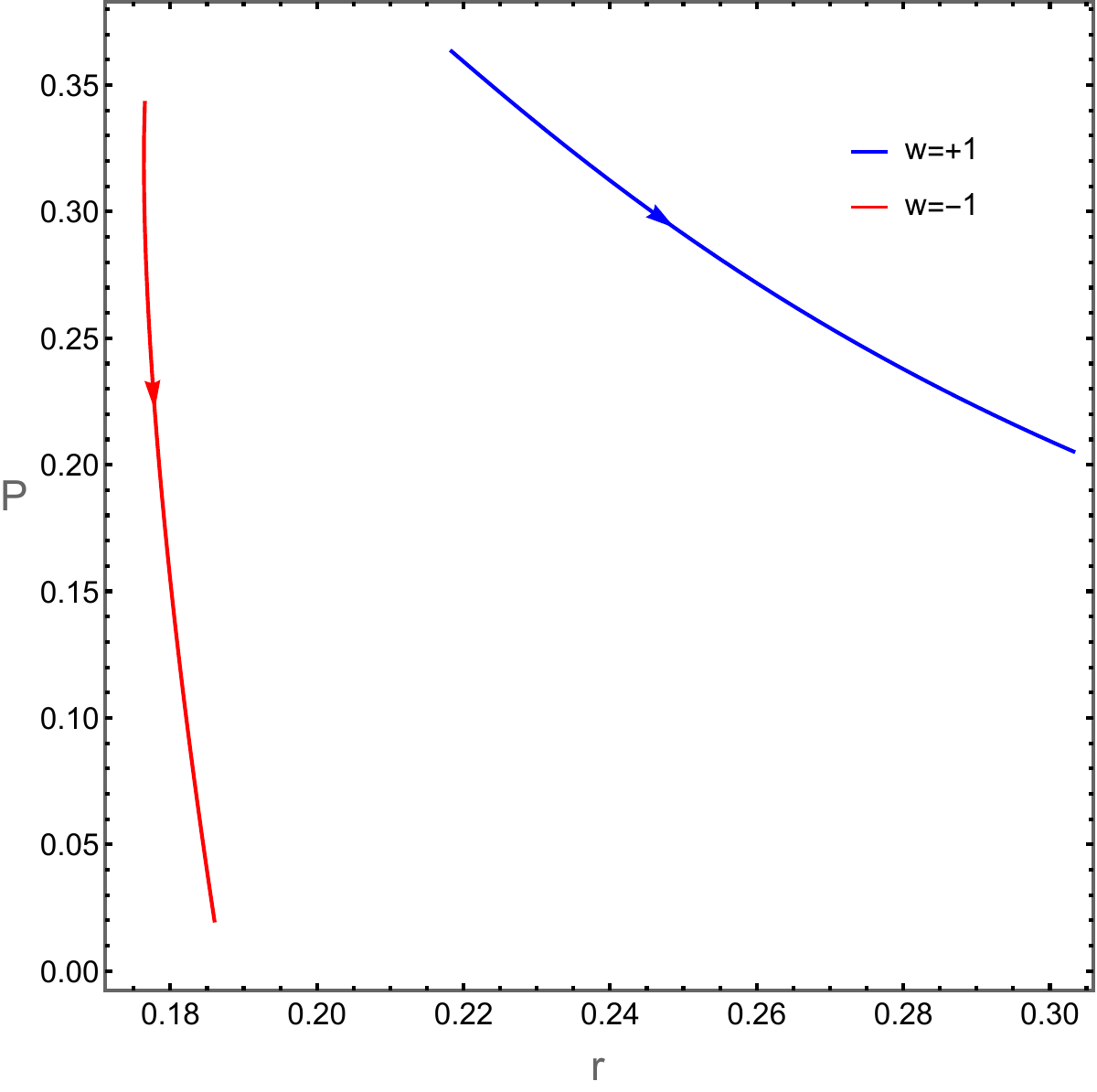}
    \caption{}
    \label{fig:trajectory}
\end{subfigure}
\hfill
\begin{subfigure}[b]{0.45\textwidth}
    \centering
    \includegraphics[width=\textwidth]{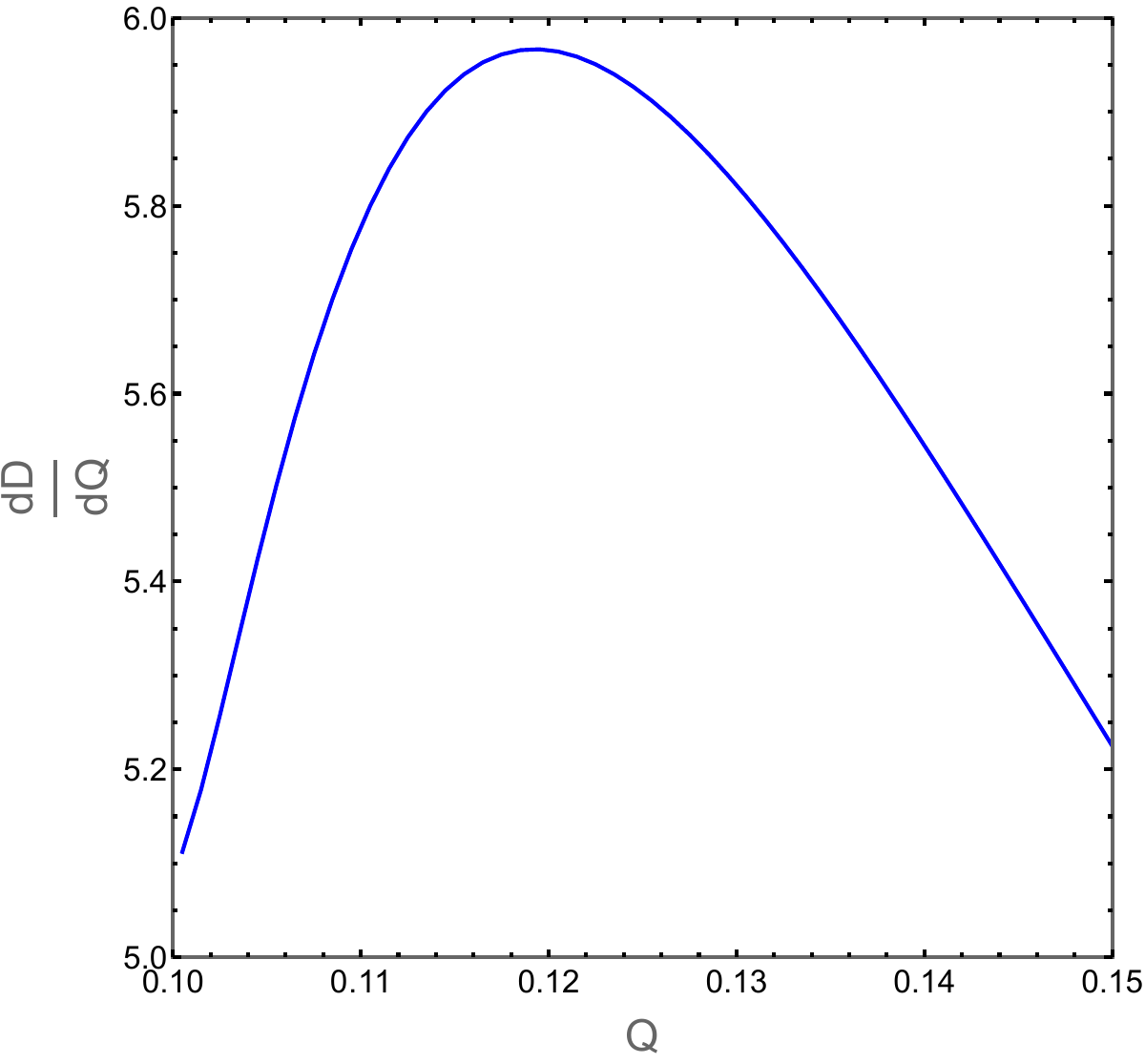}
    \caption{}
    \label{fig:separation}
\end{subfigure}

\caption{Evolution and dynamics of the thermodynamic topological defects in the Euler--Heisenberg black hole. (a) Continuous evolution of the defect positions in the thermodynamic state space.  Here we have considered $a=0.04$.}
\label{fig:defect_dynamics}
\end{figure*}

Having identified the thermodynamic defects through the winding number analysis, we now investigate how these defects evolve. For that, we use the electric charge $Q$ as control parameter. 
The numerical continuation reveals the existence of two real equilibrium branches carrying opposite winding numbers,
\begin{equation}
w=-1,\qquad w=+1,
\end{equation}
which evolve smoothly with increasing charge. Their trajectories are shown in Fig.~\ref{fig:trajectory}. The defect with negative winding number moves toward lower pressure while its radial coordinate changes only slightly, whereas the positive defect simultaneously shifts toward larger horizon radius and lower pressure. Throughout the parameter range studied, both defects remain well separated in the thermodynamic phase space and do not approach a common annihilation point.

To quantify the evolution of the topological defects, we compute the rate of change of the separation distance \(D\) between the two defects with respect to the charge parameter \(Q\) defined as : 
\begin{equation}
D(Q)=\sqrt{\left[r_2(Q)-r_1(Q)\right]^2+
\left[P_2(Q)-P_1(Q)\right]^2},
\label{eq:defect_separation}
\end{equation}

Figure~\ref{fig:separation} shows the variation of the defect velocity,  $\frac{\mathrm{d}D}{\mathrm{d}Q}$,
over the parameter interval considered.

The evolution speed initially increases with \(Q\), indicating that the defects separate more rapidly as the charge parameter grows. The velocity reaches a maximum around $Q \simeq 0.119$, 
after which it gradually decreases. This behavior demonstrates that the defect evolution is not uniform throughout the parameter space. Instead, the motion exhibits an intermediate regime in which the defects evolve most rapidly before slowing down as \(Q\) increases further.  Although the growth rate varies non monotonically with charge, it never changes sign which shows that no collision or annihilation occurs along this branch of solutions.
\begin{figure}[h!]
\centering
\includegraphics[width=0.8\textwidth]{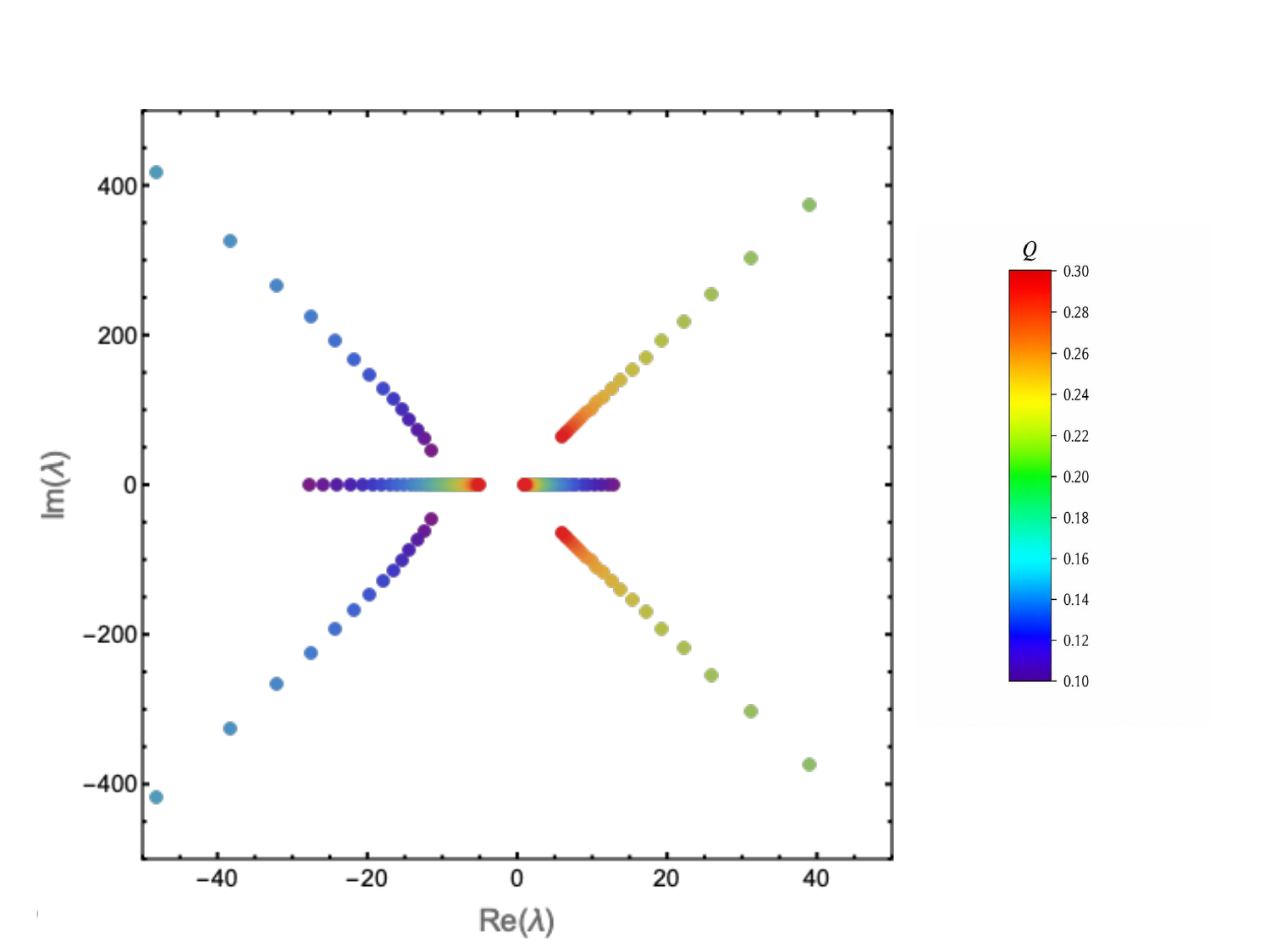}
\caption{Evolution of the Jacobian eigenvalues in the complex plane as the charge parameter $Q$ is varied.  Here $a=0.04$ is kept constant.}
\label{fig:eigen}
\end{figure}

The evolution of the Jacobian eigenvalues associated with the two topological defects is shown in Fig.~\ref{fig:eigen}. The horizontal axis represents the real part of the eigenvalues, while the vertical axis denotes their imaginary part. Each point corresponds to a particular value of the charge parameter \(Q\), with the color indicating the direction of evolution from smaller to larger values of \(Q\).

The two defects exhibit qualitatively different stability characteristics. For the stable focus defects defect carrying winding number \(w=+1\), the Jacobian possesses a pair of complex-conjugate eigenvalues,
\[
\lambda_{1,2}=\alpha \pm i\beta,
\]
which appear as two symmetric branches about the real axis.  In contrast, the defect with winding number \(w=-1\) possesses two purely real eigenvalues throughout the parameter range. Since the imaginary parts vanish identically, their trajectories lie entirely on the real axis. One eigenvalue remains positive while the other remains negative, corresponding to the two characteristic directions of a saddle point. As \(Q\) increases, both eigenvalues move smoothly toward the origin without crossing it, demonstrating that the saddle character of the defect is preserved during the evolution.
An interesting feature of Fig.~\ref{fig:eigen} is the clear geometric separation between the two classes of defects. The \(w=+1\) defect occupies the complex plane through a pair of complex-conjugate branches, whereas the \(w=-1\) defect is confined entirely to the real axis. The resulting ``X''-shaped structure together with the horizontal branches provides a compact visualization of the fundamentally different local dynamics associated with the two topological charges. Interestingly, the symmetric arrangement of the eigenvalue trajectories in the complex plane bears a qualitative resemblance to the well-known distributions of Lee--Yang zeros encountered in black hole thermodynamics.  Whether this similarities reflects a more meaningful connection between topological defect dynamics and the Lee--Yang theory of phase transitions remains an interesting question for future investigation.

Another important observation is that the eigenvalue trajectories remain well separated throughout the entire parameter interval considered. Neither the real eigenvalues of the \(w=-1\) defect nor the complex-conjugate eigenvalues of the \(w=+1\) defect undergo any collision, merging, or branch crossing in the complex plane. Consequently, there is no evidence of an eigenvalue degeneracy or exchange of stability between the two defects. Instead, the eigenvalues evolve smoothly as the charge parameter \(Q\) is varied. 

The evolution of the defect velocities is presented in \textbf{Appendix 1.} The analysis of radial velocity and pressure velocity vary continuously along each branch, indicating that the motion is smooth without any discontinuous jumps or sudden changes of direction (\textbf{See Appendix 1}) . Although the magnitudes of the velocities differ for the two branches, both defects evolve in a regular and deterministic manner as the control parameter is varied.

\section{6D Gauss Bonnet AdS Black Hole}

In this section,  we consider the charged Gauss--Bonnet--AdS black hole in a $D$-dimensional spacetime, whose line element is given by~\cite{Cai:2013qga}
\begin{align}
\label{line_element_dD}
ds^2=-f(r)\,dt^2+\frac{dr^2}{f(r)}+r^2 h_{ij}dx^i dx^j,
\end{align}
where $h_{ij}$ denotes the metric on a $(D-2)$-dimensional maximally symmetric space with constant curvature $(D-2)(D-3)k$. The parameter $k$ characterizes the horizon geometry, with $k=1$, $0$, and $-1$ corresponding to spherical, planar, and hyperbolic horizons, respectively. In the present work, we restrict our analysis to the physically relevant spherical topology ($k=1$), for which the metric function is
\begin{equation}
f(r)=1+\frac{r^2}{2\tilde{\alpha}}
\left(
1-
\sqrt{
1+\frac{64\pi M\tilde{\alpha}}{(D-2)r^{D-1}}
-\frac{2q^2\tilde{\alpha}}{(D-2)(D-3)r^{2D-4}}
+\frac{8\tilde{\alpha}\Lambda}{(D-1)(D-2)}
}
\right),
\end{equation}

where
\[
\tilde{\alpha}=(D-3)(D-4)\alpha,
\]
with $\alpha$ denoting the Gauss--Bonnet coupling constant, $M$ the black hole mass, and $q$ the electric charge. 

The event horizon is located at the largest positive root of the equation
\[
f(r_+)=0.
\]
Solving this condition for the mass parameter yields
\begin{multline}
M=
\frac{r_+^{-D-5}}
{16\pi(D-2)l^2}
\Bigl[
(D-3)(D-2)r_+^{2D}
\Bigl(
\bigl[\alpha(D-4)(D-3)+r_+^2\bigr]l^2
+r_+^4
\Bigr)
+2l^2q^2r_+^8
\Bigr],
\end{multline}
The expression for temperature is obtained to be 
\begin{equation}
T=\frac{(D-3) r^{-2 D-1} \left(l^2 \left(\left(D^2-5 D+6\right) r^{2 D} \left(\alpha  \left(D^2-9 D+20\right)+r^2\right)-2 q^2 r^8\right)+\left(D^2-3 D+2\right) r^{2 D+4}\right)}{4 \pi  (D-2)^2 l^2 \left(2 \alpha  \left(D^2-7 D+12\right)+r^2\right)}
\end{equation}
Here we substitute $D=6$ and $P= \frac{3}{8\pi l^2}$ for the rest of our analysis. \\

Again we investigate the thermodynamic topological structure of the criticalities in charged Gauss--Bonnet--AdS black hole by tracking the zeros of the vector field
\begin{equation}
\bm{\phi}(r_+,P)
=
\left(
\phi_1,\phi_2
\right)
=
\left(
\frac{1}{C_P},
\frac{\partial}{\partial r_+}\frac{1}{C_P}
\right),
\end{equation}

Ffixing the Gauss--Bonnet coupling at $\alpha=0.00805$, we find that the number of thermodynamic critical points depends sensitively on the value of the electric charge $q$. For a certain range of $q$, the system possesses three distinct critical points, whereas outside this range only a single critical point survives. The winding numbers and the corresponding eigenvalue structure of the Jacobian matrix are summarized in Table~\ref{tab:GBcritical}.

\begin{table}[ht]
\centering
\caption{Topological classification of the thermodynamic critical points for the charged Gauss--Bonnet--AdS black hole with $\alpha=0.00805$.}
\label{tab:GBcritical}
\begin{tabular}{ccccc}
\toprule
Configuration & Critical Point & $w$ & Eigenvalue Signs & Total Winding Number \\
\midrule

\multirow{3}{*}{Three critical points}
& CP$_1$ & $-1$ & $(-,+)$ & \multirow{3}{*}{$-1$} \\
& CP$_2$ & $+1$ & $(-,-)$ & \\
& CP$_3$ & $-1$ & $(-,+)$ & \\
\midrule

Single critical point
& CP & $-1$ & $(-,+)$ & $-1$ \\
\bottomrule

\end{tabular}
\end{table}

In the multicritical regime, the Gauss--Bonnet--AdS black hole admits three isolated thermodynamic defects. Two of these defects carry negative winding number ($w=-1$), while the remaining defect possesses positive winding number ($w=+1$). The Jacobian eigenvalues further distinguish their local structures. The defects with $w=-1$ exhibit one positive and one negative eigenvalue, indicating saddle-type critical points, whereas the defect with $w=+1$ possesses two negative eigenvalues, corresponding to a stable node in the thermodynamic vector field.

As the electric charge is varied, the multicritical structure continuously evolves. Beyond a critical value of the charge, only a single thermodynamic critical point remains. Since the total number of critical points changes from three to one, it is natural to ask how the missing defects disappear.

To answer this question, we track the evolution of the zeros of the thermodynamic vector field.
For a certain range of the charge parameter, three thermodynamic defects are
present. Denoting their trajectories as
\begin{equation}
\bm{x}_i(q)=\left(r_i(q),P_i(q)\right),
\qquad i=1,2,3 ,
\end{equation}
we observe that two of the defects approach each other as $q$ increases,
while the third defect moves away from the collision region. To quantify this
behavior, we define the separation between two defects as
\begin{equation}
D_{ij}(q)
=
\left|
\bm{x}_i(q)-\bm{x}_j(q)
\right|
=
\sqrt{
\left(r_i-r_j\right)^2+
\left(P_i-P_j\right)^2
}.
\end{equation}

\begin{figure}
\centering
\includegraphics[width=0.48\textwidth]{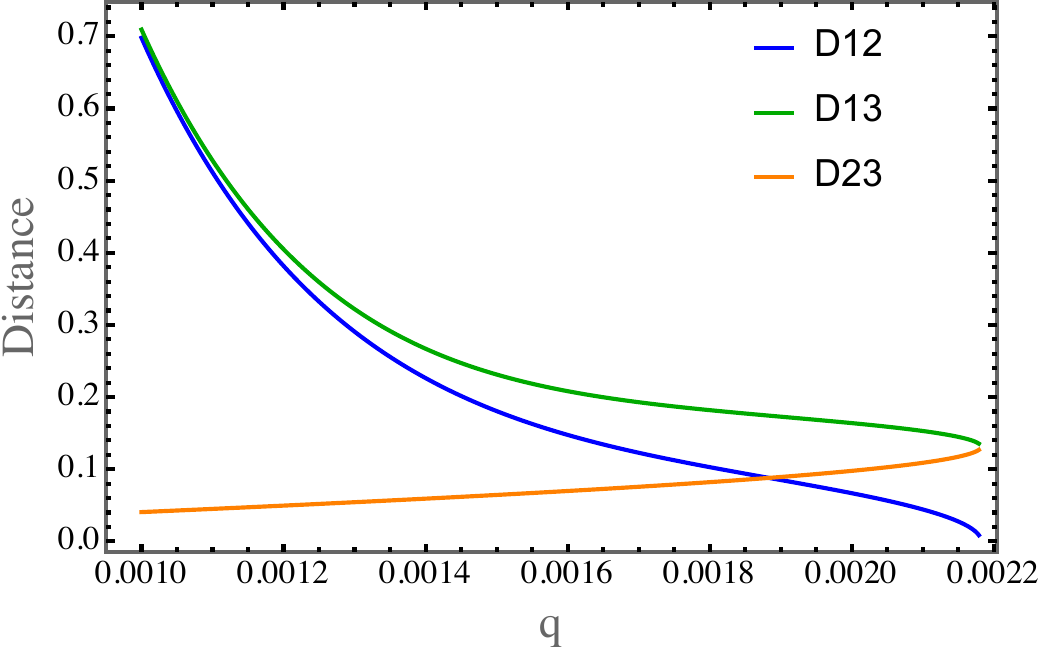}
\hfill
\includegraphics[width=0.48\textwidth]{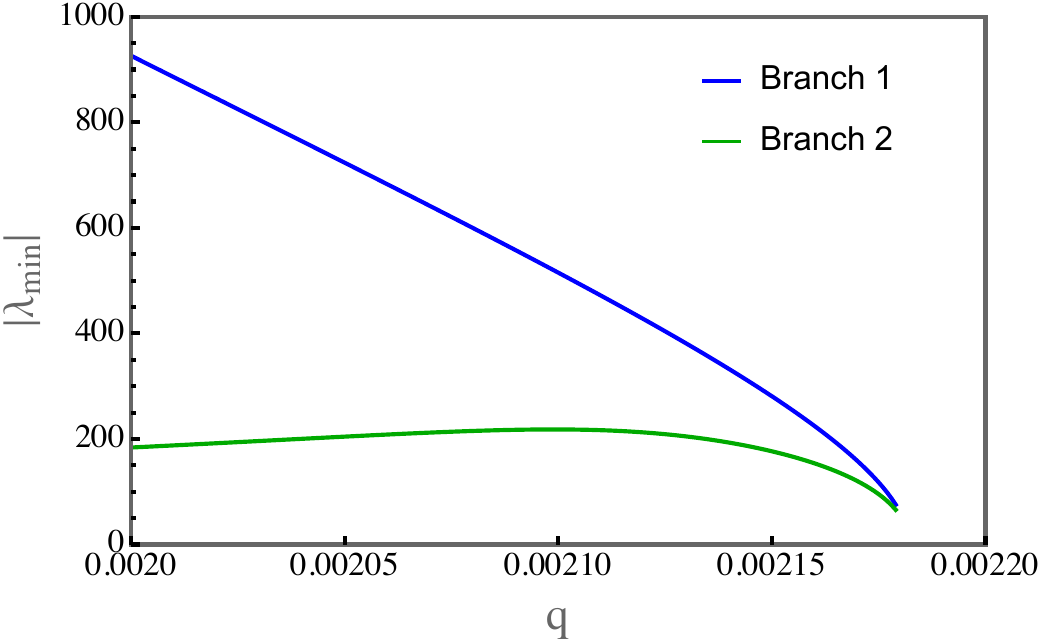}
\caption{Trajectory plot of zeros of the vector field and minimum eigen value plot at the zero points. Here $\alpha=0.00815$ kept constant.}
\label{fig:distance}
\end{figure}

The left panel of Figure~\ref{fig:distance} shows the evolution of the pairwise distances between the three thermodynamic defects as the charge parameter $q$ is varied. The blue curve, corresponding to $D_{12}$, decreases continuously and finally becomes zero at the critical charge $q=q_c$. This clearly shows that defects 1 and 2 move towards each other and eventually merge.

The green curve ($D_{13}$) also decreases with increasing charge, indicating that defect 3 moves closer to defect 1. However, the distance never becomes zero, showing that defect 3 does not collide with defect 1. On the other hand, the orange curve ($D_{23}$) increases gradually during the evolution. At the critical charge, the green and orange curves meet, i.e.,
\begin{equation}
D_{13}=D_{23}.
\end{equation}
This occurs because, at $q=q_c$, defects 1 and 2 become coincident. As a result, the remaining defect is at the same distance from both of them.

Therefore, the  analysis shows that only defects 1 and 2 participate in the collision, while the third defect remains separated. The vanishing of $D_{12}$ provides clear evidence of the defect collision, whereas the equality $D_{13}=D_{23}$ at $q=q_c$ confirms that the remaining defect is equidistant from the merged pair.

The critical collision point can be identified as a degenerate zero of the
vector field. In particular, besides satisfying
\begin{equation}
\phi_1=\phi_2=0,
\end{equation}
the Jacobian of the vector field becomes singular,
\begin{equation}
\det(J)=0,
\end{equation}
Solving these three conditions simultaneously at $\alpha=0.00815$ gives the critical collision
point
\begin{equation}
(r_c,P_c,q_c)
=
(0.215804,\;0.115339,\;0.00218216),
\end{equation}
where the two defect branches become degenerate.

The annihilation phenomenon is further confirmed by analysing the eigenvalues
of the Jacobian matrix. Let $\lambda_1$ and $\lambda_2$ denote the eigenvalues
of $J$. Since
\begin{equation}
\det(J)=\lambda_1\lambda_2,
\end{equation}
the degeneracy of the Jacobian implies that at the collision point at least
one eigenvalue approaches zero. In the right panel of Fig.\ref{fig:distance}, the numerical evolution of the smallest eigenvalue magnitude  along the two approaching branches shows
\begin{equation}
|\lambda_{\rm min}^{(1)}|
\rightarrow0,
\qquad
|\lambda_{\rm min}^{(2)}|
\rightarrow0,
\end{equation}
at the same value of $q=q_c$. This demonstrates that the two isolated zeros
of the vector field lose their individual identity and transform into a
degenerate critical configuration.

\begin{figure*}[h!]
\begin{subfigure}{0.32\textwidth}
    \centering
    \includegraphics[width=\textwidth]{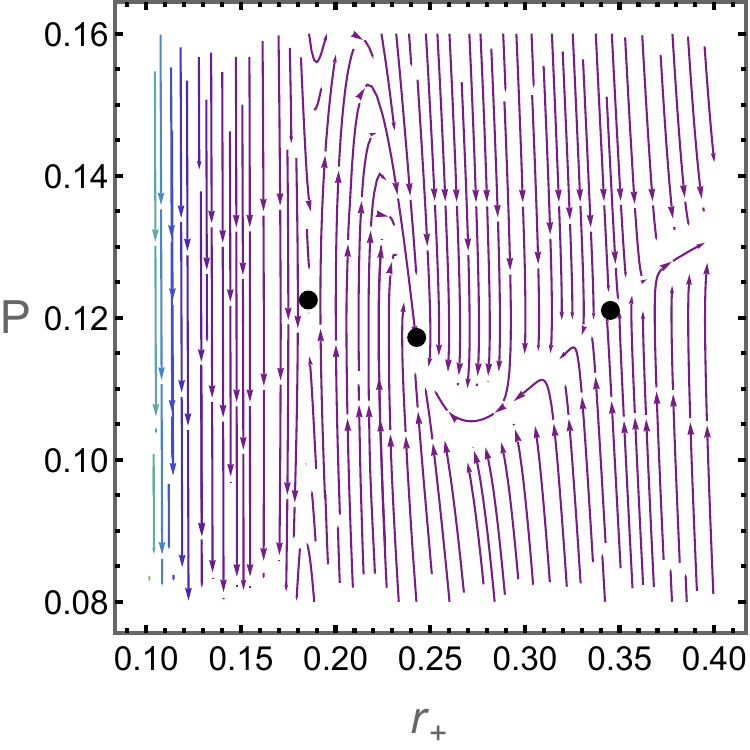}
    \caption{Before collision}
    \label{}
\end{subfigure}
\hfill
\begin{subfigure}[b]{0.32\textwidth}
    \centering
    \includegraphics[width=\textwidth]{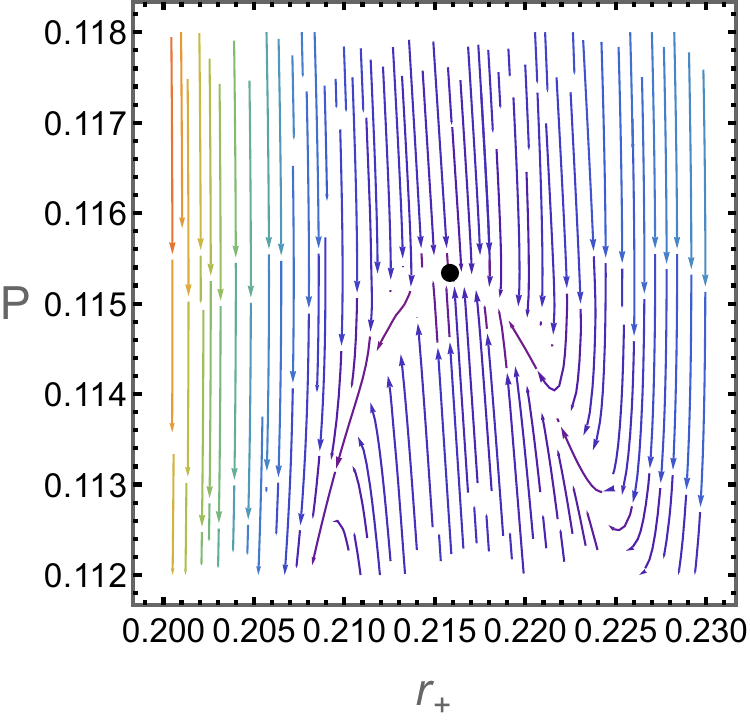}
    \caption{During collision}
    \label{}
\end{subfigure}
\hfill
\begin{subfigure}[b]{0.32\textwidth}
    \centering
    \includegraphics[width=\textwidth]{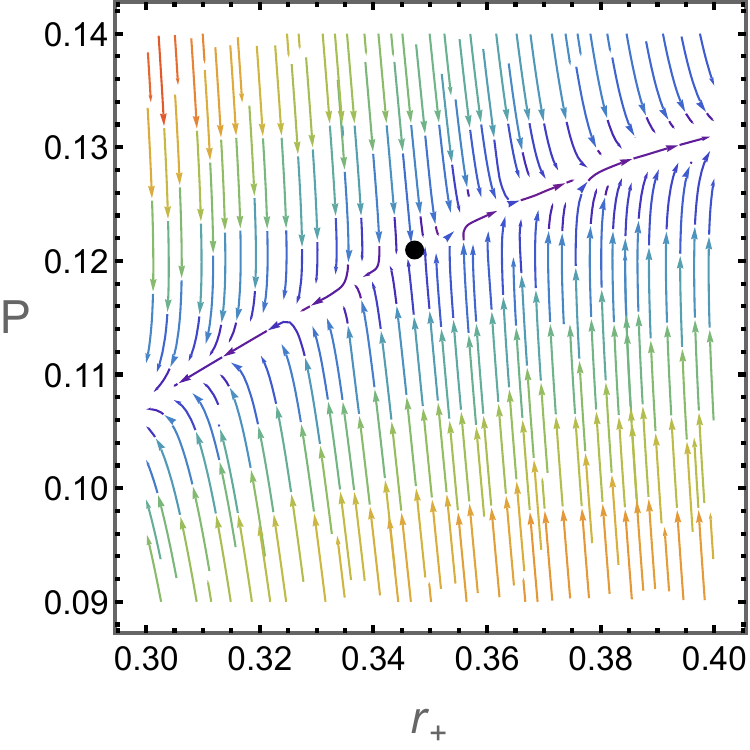}
    \caption{After collision}
    \label{}
\end{subfigure}

\caption{Annihilation of the thermodynamic topological defects in the 6D Gaus Bonnet black hole. for $q=0.00205$ and $\alpha=0.00815$ }
\label{col}
\end{figure*}

Therefore, the defect evolution can be summarized as
\begin{equation}
\bm{x}_1(q)+\bm{x}_2(q)
\longrightarrow
\bm{x}_c,
\end{equation}
followed by the disappearance of the pair for $q>q_c$. Since the total
topological charge must be conserved, such an annihilation process requires
the two merging defects to carry opposite winding numbers,
\begin{equation}
w_1+w_2=0.
\end{equation}
Hence, the charge parameter $q$ drives a genuine topological transition in
the thermodynamic phase space, where a pair of defects with opposite
topological charges annihilate while the remaining defect persists. The vector fields in Fig. \ref{col} represent the same annihilation phenomena through streamline plots. \\

\subsection{Einstein-Power-Maxwell black holes and multicriticality}
Next,  we consider the charged AdS black hole solution in generalized Power-Maxwell electrodynamics introduced in Ref.~\cite{PowerMaxwell}. In this theory, the electromagnetic sector is extended by including higher-order powers of the Maxwell invariant, each accompanied by an independent coupling constant $\alpha_n$.  \\
The general form of the action for non-linear electrodynamics minimally coupled to $D=4$
Einstein gravity  is 
\begin{equation}
S=\int d^{4}x\,\sqrt{-g}\left(
R-2\Lambda-\sum_{i=1}^{N}\alpha_i\left(F^{2}\right)^i
\right),
\end{equation}
where, \[
F^{2}\equiv F_{\mu\nu}F^{\mu\nu},
\qquad
F_{\mu\nu}\equiv \nabla_{\mu}A_{\nu}-\nabla_{\nu}A_{\mu},
\]

Here $R$ is the Ricci scalar. The  dimensional coupling constants $\alpha_i$ are expressed as  $[\alpha_i]=L^{2(i-1)} $.  $A_\mu$ is the $U(1)$ Maxwell field.  The Einstein-Maxwell theory is recovered when $\alpha_1=1$ and $\alpha_i (i>1)=0$.\\
Using the ansatz
\begin{align}
ds^2 &= -U(r)\,dt^2 + \frac{1}{U(r)}\,dr^2 + r^2 d\Omega_2^2, \\
A_\mu &= \left(\Phi(r),\,0,\,0,\,0\right).
\end{align}
The metric function is obtained as
\begin{equation}
U
= 1 - \frac{2M}{r}
+ \frac{Q^2}{r^2}
+ \frac{b_5 Q}{2r^6}
+ \frac{b_9 Q}{3r^{10}}
+ \frac{b_{13}Q}{4r^{14}}
+ \frac{b_{17}Q}{5r^{18}}
+ \frac{b_{21}Q}{6r^{22}}
+ \frac{b_{25}Q}{7r^{26}}
+ \frac{r^2}{l^2}.
\end{equation}
For the detail derivation of the black hole solution one can be referred to Ref.~\cite{PowerMaxwell}.
The relevant thermodynamic quantities are calculated to be
\begin{align}
M
&= \frac{1}{840\,l^2 r^{25}}
\Big(
60 b_{25} l^2 Q
+ 70 b_{21} l^2 Q r^4
+ 84 b_{17} l^2 Q r^8
+ 105 b_{13} l^2 Q r^{12}
\notag\\
&\qquad
+ 140 b_{9} l^2 Q r^{16}
+ 210 b_{5} l^2 Q r^{20}
+ 420 l^2 Q^2 r^{24}
+ 420 l^2 r^{26}
+ 420 r^{28}
\Big).\\[1ex]
T &= \frac{1}{4\pi r_+}
\Bigg(
1 + \frac{3r_+^2}{l^2}
- \frac{Q^2}{r_+^2}
- \frac{5b_5Q}{2r_+^6}
- \frac{3b_9Q}{r_+^{10}}
\nonumber\\
&\qquad
- \frac{13b_{13}Q}{4r_+^{14}}
- \frac{17b_{17}Q}{5r_+^{18}}
- \frac{7b_{21}Q}{2r_+^{22}}
- \frac{25b_{25}Q}{7r_+^{26}}
\Bigg),\\[1ex]
\end{align}
Here we substitute $P= \frac{3}{8\pi l^2}$ for the rest of our analysis. Next we calculate our proposed vector field components as
\begin{equation}
\bm{\phi}(r,P)=
\left(
\frac{1}{C_P},
\frac{\partial}{\partial r}\left(\frac{1}{C_P}\right)
\right).
\end{equation}
where $C_P=\frac{dM}{dT}$.  For our analysis, the coupling constants are considered to be :
\begin{align}
b_{1} &= Q,\quad b_{5} = \frac{4}{5}Q^{3}\alpha_{2}, \quad b_{9} = \frac{4}{3}Q^{5}\left(4\alpha_{2}^{2}-\alpha_{3}\right),\\[0.5ex]
b_{13} &= \frac{32}{13}Q^{7}\left(
24\alpha_{2}^{3}
-12\alpha_{3}\alpha_{2}
+\alpha_{4}
\right),\\[0.5ex]
b_{17} &= \frac{80}{17}Q^{9}\left(
176\alpha_{2}^{4}
-132\alpha_{2}^{2}\alpha_{3}
+16\alpha_{4}\alpha_{2}
+9\alpha_{3}^{2}
-\alpha_{5}
\right),\\[0.5ex]
b_{21} &= \frac{64}{7}Q^{11}\left(
1456\alpha_{2}^{5}
+234\alpha_{3}^{2}\alpha_{2}
+208\alpha_{4}\alpha_{2}^{2}
-24\alpha_{4}\alpha_{3}
-1456\alpha_{2}^{3}\alpha_{3}
-20\alpha_{5}\alpha_{2}
+\alpha_{6}
\right),\\[0.5ex]
b_{25} &= \frac{448}{25}Q^{13}\left(
13056\alpha_{2}^{6}
+2560\alpha_{2}^{3}\alpha_{4}
-720\alpha_{3}\alpha_{2}\alpha_{4}
+16\alpha_{4}^{2}
-300\alpha_{2}^{2}\alpha_{5}
\right. \notag\\
&\qquad\left.
+4320\alpha_{2}^{2}\alpha_{3}^{2}
-16320\alpha_{2}^{4}\alpha_{3}
+24\alpha_{6}\alpha_{2}
-135\alpha_{3}^{3}
+30\alpha_{3}\alpha_{5}
-\alpha_{7}
\right).
\end{align}

For the numerical analysis presented in this section, we fix the electric charge and higher-order coupling constants as
\begin{align}
Q &= 6.623, \qquad \alpha_{1}=1,\nonumber\\
\alpha_{2} &= -21.63694203,\qquad
\alpha_{3}=1493.535254,\qquad
\alpha_{4}=-148046.3896,\nonumber\\
\alpha_{5} &=1.759261993\times10^{7},\qquad
\alpha_{6}=-2.332423991\times10^{9},\qquad
\alpha_{7}=3.327781293\times10^{11}.
\end{align}

\begin{table}[t!]
\centering
\caption{Topological classification of the thermodynamic critical points for the Power-Maxwell black hole. As the charge parameter $Q$ increases, the system evolves through the sequence $2\rightarrow4\rightarrow6$, while the total winding number remains conserved.}
\label{tab:PMcritical}
\begin{tabular}{ccccc}
\toprule
Configuration & Critical Point & $w$ & Eigenvalue Signs & $W_{\rm tot}$ \\
\midrule

\multirow{2}{*}{Two critical points}
& CP$_1$ & $-1$ & $(-,+)$ & \multirow{2}{*}{$-2$} \\
& CP$_2$ & $-1$ & $(-,+)$ & \\
\midrule

\multirow{4}{*}{Four critical points}
& CP$_1$ & $-1$ & $(-,+)$ & \multirow{4}{*}{$-2$} \\
& CP$_2$ & $-1$ & $(-,+)$ & \\
& CP$_3$ & $+1$ & $(-,-)$ & \\
& CP$_4$ & $-1$ & $(-,+)$ & \\
\midrule

\multirow{6}{*}{Six critical points}
& CP$_1$ & $-1$ & $(-,+)$ & \multirow{6}{*}{$-2$} \\
& CP$_2$ & $-1$ & $(-,+)$ & \\
& CP$_3$ & $+1$ & $(-,-)$ & \\
& CP$_4$ & $-1$ & $(-,+)$ & \\
& CP$_5$ & $+1$ & $(-,-)$ & \\
& CP$_6$ & $-1$ & $(-,+)$ & \\
\bottomrule
\end{tabular}
\end{table}

The number of thermodynamic topological defects of the Power-Maxwell black hole depends sensitively on the charge parameter $Q$. By analyzing the zero points of the constructed thermodynamic vector field, we find that the system undergoes a sequence of topological transitions as the charge is varied. For relatively small values of $Q$, the thermodynamic state space contains two topological defects. As the charge increases, a pair of defects with opposite Brouwer degree is created through a saddle-node bifurcation, giving rise to an intermediate configuration with four critical points. Upon further increasing $Q$, another neutral pair of defects is generated, resulting in a six-defect configuration. Consequently, the evolution of the thermodynamic defects follows the sequence
\begin{equation}
2 \longrightarrow 4 \longrightarrow 6,
\end{equation}
while the total winding number remains conserved,
\begin{equation}
W_{\rm tot}=-2,
\end{equation}
throughout the entire evolution. The conservation of the total topological charge demonstrates that new defects are always created in pairs carrying opposite Brouwer degrees, so that each bifurcation leaves the net winding number unchanged.

To characterize the multicritical points locally, we investigate the Brouwer degree of each topological defect. The Brouwer degree, determined from the sign of the Jacobian determinant, specifies the topological charge of each defect, whereas the eigenvalues of the Jacobian characterize the local stability of the thermodynamic flow.  In the two-defect configuration, both critical points possess negative Brouwer degree ($\omega=-1$), corresponding to saddle-type defects. After the first bifurcation, the newly created defect pair consists of one positive and one negative topological charge, producing a four-defect configuration with three saddle points ($\omega=-1$) and one stable node ($\omega=+1$). A second pair-creation event generates another pair with opposite Brouwer degrees, leading to the six-defect configuration, which contains four saddle defects and two stable nodes. The local vector fields are fully consistent with this classification and provide a direct visualization of the thermodynamic flow around each defect.

Although defects carrying the same Brouwer degree possess identical topological charge, their Jacobian eigenvalues reveal different local flow geometries and stability strengths. The locations of the critical points together with their Brouwer degrees and Jacobian eigenvalue signatures are summarized in Table~\ref{tab:PMcritical}.

\section{Conclusion and Discussions}
In this work, we proposed a new topological framework based on the heat capacity to study thermodynamic multicritical points of black holes. By constructing a two-dimensional thermodynamic vector field, we showed that the zeros of the vector field correspond to the multicritical points of the system. The Brouwer degree of each zero provides a topological classification of the corresponding critical point, while the Jacobian eigenvalues determine its local stability.

We applied this method to several AdS black hole solutions, namely the RN--AdS, Euler--Heisenberg--AdS, six-dimensional Gauss--Bonnet--AdS, and Power--Maxwell--AdS black holes. Our analysis shows that although the number of thermodynamic critical points changes as the thermodynamic parameters are varied, the total topological number of each black hole system remains conserved. The appearance of new critical points occurs through the creation of topologically neutral defect pairs, leaving the total topological charge unchanged.  Different black hole systems can be characterized by different conserved total topological numbers. For example, the RN--AdS and six-dimensional Gauss--Bonnet--AdS black holes have total winding number $W_{\rm tot}=-1$, the Euler--Heisenberg--AdS black hole has $W_{\rm tot}=0$, while the Power--Maxwell--AdS black hole has $W_{\rm tot}=-2$. 

The main results of the present work are summarized in Table~\ref{tab:summary}. 

\begin{table}[ht]
\centering
\caption{Summary of the topology of thermodynamic multi-critical points for the black hole solutions studied in this work.}
\label{tab:summary}
\begin{tabular}{ccc}
\toprule
Black Hole & Number of Critical Points & Total Topological Number ($W_{\rm tot}$) \\
\midrule
RN--AdS & $1,\;0$ & $-1$ \\

Euler--Heisenberg--AdS & $2,\;0$ & $0$ \\

6D Gauss--Bonnet--AdS & $3,\;1$ & $-1$ \\

Power--Maxwell--AdS & $2,\;4,\;6$ & $-2$ \\
\bottomrule
\end{tabular}
\end{table}
From a physical point of view, our results suggest that the total topological number acts as a global invariant that characterizes the multicritical structure of a black hole.  The conservation of the total topological charge indicates that the overall structure of the thermodynamic phase space is preserved, and new critical points can only appear or disappear through the creation or annihilation of topologically neutral defect pairs.  The Brouwer degree and the Jacobian eigenvalues determine the local behavior of the thermodynamic flow around it. In particular, saddle points correspond to unstable directions in the thermodynamic phase space, whereas stable nodes and stable spirals represent locally attracting configurations. Together, the Brouwer degree and the Jacobian eigenvalues provide both a global topological and a local dynamical description of thermodynamic multicriticality.

In our next work, we plan to extend the present framework to condensed matter systems that exhibit multicritical behavior. Since multicriticality is well established in several condensed matter models, applying our topological construction to these systems may provide a clearer physical interpretation of the conserved total winding number and the evolution of topological defects. A comparison between condensed matter systems and their black hole counterparts may also show similar topological features shared by these seemingly different physical systems.
\section{Appendix}

\begin{figure}[t!]
\centering
\includegraphics[width=0.45\textwidth]{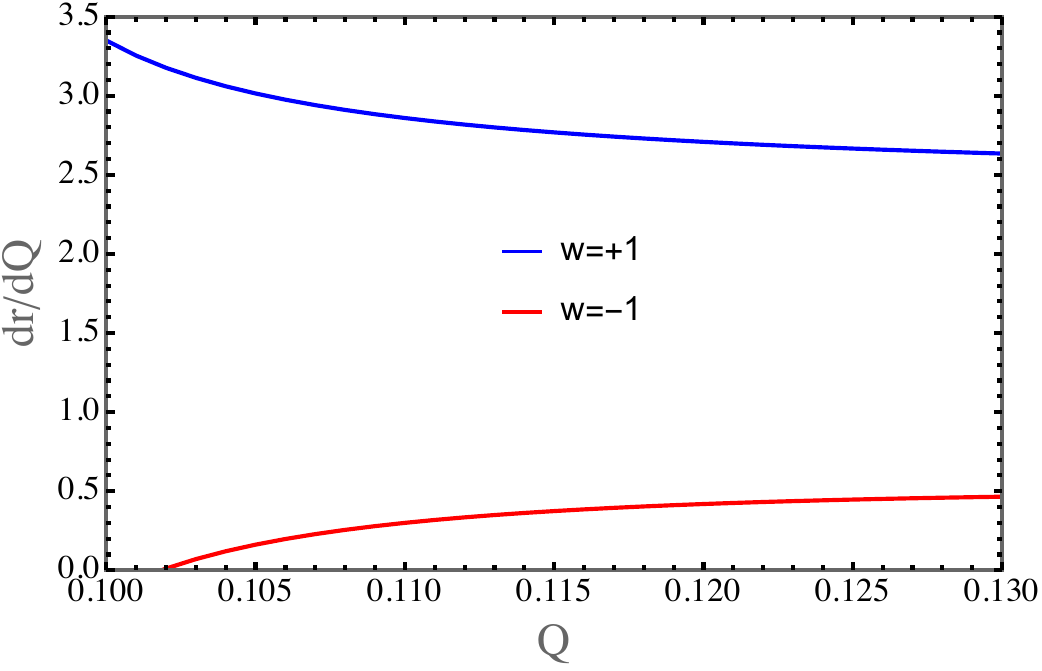}
\hfill
\includegraphics[width=0.45\textwidth]{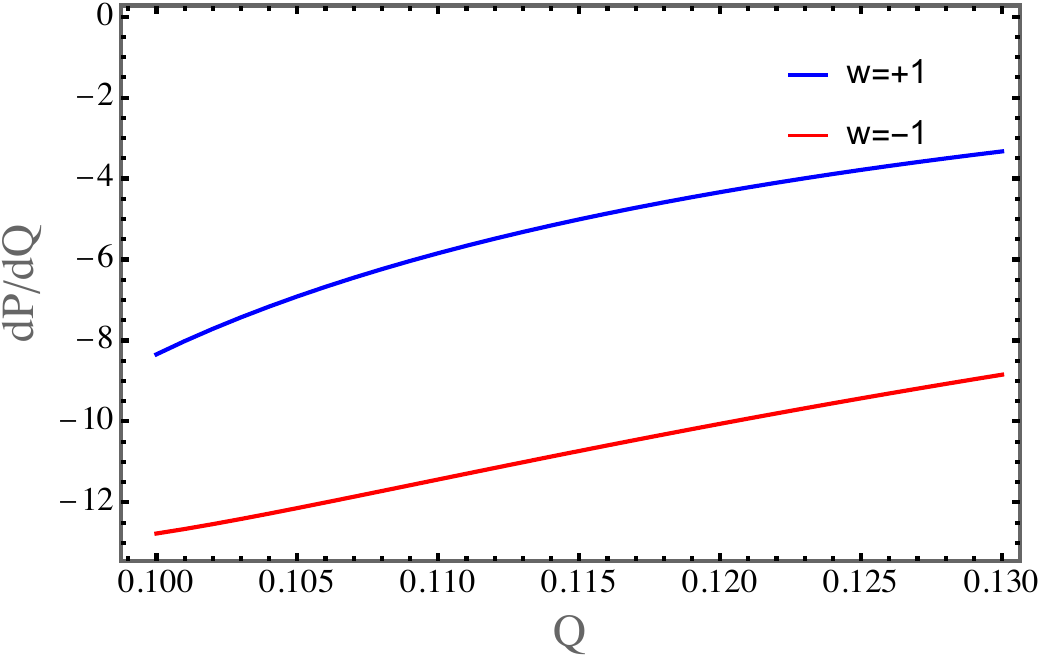}
\caption{Velocity of the topological defect for Euler Heisenberg black hole at $a=0.04$}
\label{fig:velocity}
\end{figure}

In this section, we investigate the evolution of the thermodynamic defect velocities in the Euler--Heisenberg black hole as the charge parameter is varied.
We regard their locations as functions of $Q$,
\begin{equation}
r_c=r_c(Q), \qquad P_c=P_c(Q).
\end{equation}
Since the critical points are defined as the simultaneous zeros of the
thermodynamic vector field,
\begin{equation}
\bm{\phi}(r_c(Q),P_c(Q),Q)=0,
\label{zerocondition}
\end{equation}
the above relation must remain satisfied for every value of $Q$.

Differentiating Eq.~(\ref{zerocondition}) with respect to $Q$ and applying
the multivariable chain rule gives
\begin{align}
\frac{d\phi_1}{dQ}
&=
\frac{\partial\phi_1}{\partial r}\frac{dr_c}{dQ}
+
\frac{\partial\phi_1}{\partial P}\frac{dP_c}{dQ}
+
\frac{\partial\phi_1}{\partial Q}
=0,
\\
\frac{d\phi_2}{dQ}
&=
\frac{\partial\phi_2}{\partial r}\frac{dr_c}{dQ}
+
\frac{\partial\phi_2}{\partial P}\frac{dP_c}{dQ}
+
\frac{\partial\phi_2}{\partial Q}
=0.
\end{align}

These two equations can be expressed in matrix form as
\begin{equation}
\begin{pmatrix}
\dfrac{\partial\phi_1}{\partial r} &
\dfrac{\partial\phi_1}{\partial P}
\\[2mm]
\dfrac{\partial\phi_2}{\partial r} &
\dfrac{\partial\phi_2}{\partial P}
\end{pmatrix}
\begin{pmatrix}
\dfrac{dr_c}{dQ}
\\[2mm]
\dfrac{dP_c}{dQ}
\end{pmatrix}
+
\begin{pmatrix}
\dfrac{\partial\phi_1}{\partial Q}
\\[2mm]
\dfrac{\partial\phi_2}{\partial Q}
\end{pmatrix}
=
\mathbf{0}.
\end{equation}

Introducing the Jacobian matrix of the thermodynamic vector field,
\begin{equation}
J=
\begin{pmatrix}
\dfrac{\partial\phi_1}{\partial r} &
\dfrac{\partial\phi_1}{\partial P}\\[2mm]
\dfrac{\partial\phi_2}{\partial r} &
\dfrac{\partial\phi_2}{\partial P}
\end{pmatrix},
\end{equation}
the above equation can be written compactly as
\begin{equation}
J
\begin{pmatrix}
\dfrac{dr_c}{dQ}\\[2mm]
\dfrac{dP_c}{dQ}
\end{pmatrix}
+
\frac{\partial\bm{\phi}}{\partial Q}
=0.
\end{equation}
Assuming that the Jacobian is nonsingular ($\det J\neq0$), the evolution of
the critical point is obtained by solving the above linear system,
\begin{equation}
\boxed{
\mathbf{V}
=
\begin{pmatrix}
\dfrac{dr_c}{dQ}\\[2mm]
\dfrac{dP_c}{dQ}
\end{pmatrix}
=
-
J^{-1}
\frac{\partial\bm{\phi}}{\partial Q}.
}
\label{velocity}
\end{equation}
Equation~(\ref{velocity}) defines the \emph{parameter-space velocity} of the
thermodynamic critical point.

Figure~\ref{fig:velocity} shows the evolution of the two thermodynamic defects of the Euler--Heisenberg black hole as the charge parameter $Q$ is varied. The upper panel displays the radial velocity $dr/dQ$ of the defects, while the lower panel shows their pressure velocity $dP/dQ$. The defect with positive Brouwer degree ($w=+1$) moves toward larger horizon radius much faster than the saddle defect ($w=-1$), as indicated by its consistently larger positive value of $dr/dQ$. As $Q$ increases, the radial velocity of the stable defect decreases gradually, whereas that of the saddle defect increases slightly, indicating that both defects continue to move outward but at different rates.

The lower panel shows that the pressure velocities of both defects remain negative throughout the evolution. This implies that, as the charge increases, both critical points continuously drift toward lower pressures in the thermodynamic phase space. Moreover, the stable defect ($w=+1$) exhibits a smaller magnitude of $dP/dq$ than the saddle defect ($w=-1$), indicating that its critical pressure decreases more slowly. The smooth evolution of both $dr/dQ$ and $dP/dQ$ demonstrates that the thermodynamic defects evolve continuously with the charge parameter. Interestingly, our numerical analysis shows that the two defects never approach one another in the thermodynamic phase space. Instead, they remain spatially separated and eventually disappear from the physical parameter space beyond a critical value of $Q$, where no positive real critical points exist.


\begin{thebibliography}{99}
		
		
	\def\EPJC{Eur. Phys. J. C\,}
\def\IJMPA{Int. J. Mod. Phys. A\,}
\def\JCAP{J. Cosmol. Astropart. Phys.\,}
\def\JHEP{J. High Energy Phys.\,}
\def\CQG{Classical Quantum Gravity\,}
\def\JMP{J. Math. Phys. (N.Y.)\,}
\def\NPB{Nucl. Phys. B \,}
\def\PDU{Phys. Dark Univ.\,}
\def\PLB{Phys. Lett. B \,}
\def\PRD{Phys. Rev. D\,}
\def\PRL{Phys. Rev. Lett.\,}
\def\PRR{Phys. Rev. Res.\,}
\def\GRG{Gen. Relativ. Gravit.\,}

	
		\bibitem{Bekenstein:1973ur}
		J.~D.~Bekenstein,
		Black holes and entropy,
		Phys. Rev. D \textbf{7}, 2333-2346 (1973)
		doi:10.1103/PhysRevD.7.2333

		\bibitem{Hawking:1974rv}
		S.~W.~Hawking,
		Black hole explosions,
		Nature \textbf{248}, 30-31 (1974)
		doi:10.1038/248030a0
		\bibitem{Hawking:1975vcx}
		S.~W.~Hawking,
		Particle Creation by Black Holes,
		Commun. Math. Phys. \textbf{43}, 199-220 (1975)
		[erratum: Commun. Math. Phys. \textbf{46}, 206 (1976)]
		doi:10.1007/BF02345020
		
		
		
		
		
		
		
		
		\bibitem{Bardeen:1973gs}
		J.~M.~Bardeen, B.~Carter and S.~W.~Hawking,
		The Four laws of black hole mechanics,
		Commun. Math. Phys. \textbf{31}, 161-170 (1973)
		doi:10.1007/BF01645742
		
		
		
		
		
		
	
	
		\bibitem{Davies:1989ey}
		P.~C.~W.~Davies,
		Thermodynamic Phase Transitions of {Kerr-Newman} Black Holes in De Sitter Space,
		Class. Quant. Grav. \textbf{6}, 1909 (1989)
		doi:10.1088/0264-9381/6/12/018
		
		
		
		
		
		
		
		
		\bibitem{Hawking:1982dh}
		S.~W.~Hawking and D.~N.~Page,
		Thermodynamics of Black Holes in anti-De Sitter Space,
		Commun. Math. Phys. \textbf{87}, 577 (1983)
		doi:10.1007/BF01208266
		
		
		
		
		
		
	
		\bibitem{Dolan:2010ha}
		B.~P.~Dolan,
		The cosmological constant and the black hole equation of state,
		Class. Quant. Grav. \textbf{28}, 125020 (2011)
		doi:10.1088/0264-9381/28/12/125020
		[arXiv:1008.5023 [gr-qc]].
	
		\bibitem{Kubiznak:2012wp}
		D.~Kubiznak and R.~B.~Mann,
		P-V criticality of charged AdS black holes,
		JHEP \textbf{07}, 033 (2012)
		doi:10.1007/JHEP07(2012)033
		[arXiv:1205.0559 [hep-th]].
		\bibitem{Kubiznak:2016qmn}
		D.~Kubiznak, R.~B.~Mann and M.~Teo,
		Black hole chemistry: thermodynamics with Lambda,
		Class. Quant. Grav. \textbf{34}, no.6, 063001 (2017)
		doi:10.1088/1361-6382/aa5c69
		[arXiv:1608.06147 [hep-th]].
			\bibitem{rp1} J.~Sadeghi, M.~Shokri, S.~Gashti Noori and M.~R.~Alipour,
RPS thermodynamics of Taub\textendash{}NUT AdS black holes in the presence of central charge and the weak gravity conjecture.
		Gen. Rel. Grav. \textbf{54} (2022)
		\bibitem{rp2}
		Y.~Ladghami, B.~Asfour, A.~Bouali, A.~Errahmani and T.~Ouali,
4D-EGB black holes in RPS thermodynamics,
		Phys. Dark Univ. \textbf{41} (2023),
		\bibitem{rp3}
		X.~Kong, Z.~Zhang and L.~Zhao,
		Restricted phase space thermodynamics of charged AdS black holes in conformal gravity,
		Chin. Phys. C \textbf{47} (2023)
		\bibitem{rp4}
		M.~R.~Alipour, J.~Sadeghi and M.~Shokri,
		WGC and WCCC of black holes with quintessence and cloud strings in RPS space,
		Nucl. Phys. B \textbf{990} (2023)
		\bibitem{rp5}
		T.~Wang and L.~Zhao,
		Black hole thermodynamics is extensive with variable Newton constant,
		Phys. Lett. B \textbf{827} (2022)
		\bibitem{rp6}
		G.~Zeyuan and L.~Zhao,
		Restricted phase space thermodynamics for AdS black holes via holography,
		Class. Quant. Grav. \textbf{39} (2022)
		\bibitem{rp7}
		Z.~Gao, X.~Kong and L.~Zhao,
	Thermodynamics of Kerr-AdS black holes in the restricted phase space,
		Eur. Phys. J. C \textbf{82} (2022)
		\bibitem{rp8}
		S.~Dutta and G.~S.~Punia,
		String theory corrections to holographic black hole chemistry,
		Phys. Rev. D \textbf{106}, no.2, 026003 (2022)
		\bibitem{rp9}
		T.~F.~Gong, J.~Jiang and M.~Zhang,
		Holographic thermodynamics of rotating black holes,
		JHEP \textbf{06}, 105 (2023)
		\bibitem{rp10}
		W.~Cong, D.~Kubiznak, R.~B.~Mann and M.~R.~Visser,
		Holographic CFT phase transitions and criticality for charged AdS black holes,
		JHEP \textbf{08}, 174 (2022)
		\bibitem{rp11}
		M.~R.~Visser,
		Holographic thermodynamics requires a chemical potential for color,
		Phys. Rev. D \textbf{105}, no.10, 106014 (2022)
		
		
		
		
		
		
	

\bibitem{Altamirano:2013uqa}
N.~Altamirano, D.~Kubiznak, R.~B.~Mann and Z.~Sherkatghanad,
Kerr-AdS analogue of triple point and solid/liquid/gas phase transition,
\emph{Class. Quant. Grav.} \textbf{31}, 042001 (2014),
doi:10.1088/0264-9381/31/4/042001
[arXiv:1308.2672 [hep-th]].

\bibitem{Hennigar:2017apu}
R.~A.~Hennigar, R.~B.~Mann and E.~Tjoa,
Superfluid Black Holes,
\emph{Phys. Rev. Lett.} \textbf{118}, 021301 (2017),
doi:10.1103/PhysRevLett.118.021301
[arXiv:1609.02564 [hep-th]].


\bibitem{t1}
	S.~W.~Wei and Y.~X.~Liu,
	Topology of black hole thermodynamics,
	Phys. Rev. D \textbf{105}, no.10, 104003 (2022)
\bibitem{t2}
S.~W.~Wei, Y.~X.~Liu and R.~B.~Mann, Black Hole Solutions as Topological Thermodynamic Defects,
Phys. Rev. Lett. \textbf{129}, no.19, 191101 (2022)
\bibitem{t3}
D.~Wu, W.~Liu, S.~Q.~Wu and R.~B.~Mann, Novel topological classes in black hole thermodynamics,
Phys. Rev. D \textbf{111}, no.6, L061501 (2025)

\bibitem{t4}
B.~Hazarika, N.~J.~Gogoi and P.~Phukon,  Revisiting thermodynamic topology of Hawking-Page and Davies type phase transitions, JHEAp \textbf{45}, 87-95 (2025)


\bibitem{t5}
P.~K.~Yerra, C.~Bhamidipati and S.~Mukherji, Topology of critical points and Hawking-Page transition,
Phys. Rev. D \textbf{106}, no.6, 064059 (2022)

\bibitem{t6}
P.~K.~Yerra, C.~Bhamidipati and S.~Mukherji,  Topology of critical points in boundary matrix duals,
JHEP \textbf{03}, 138 (2024)

\bibitem{PRL119-251102}
P.V.P. Cunha, E. Berti, and C.A.R. Herdeiro,
Light Ring Stability in Ultra-Compact Objects,
\href{http://dx.doi.org/10.1103/PhysRevLett.119.251102}
{\PRL \textbf{119}, 251102 (2017)}.

\bibitem{PRL124-181101}
P.V.P. Cunha, and C.A.R. Herdeiro,
Stationary Black Holes and Light Rings,
\href{http://dx.doi.org/10.1103/PhysRevLett.124.181101}
{\PRL \textbf{124}, 181101 (2020)}.

\bibitem{PRD102-064039}
S.-W. Wei,
Topological charge and black hole photon spheres,
\href{https://doi.org/10.1103/PhysRevD.102.064039}
{\PRD \textbf{102}, 064039 (2020)}.

\bibitem{PRD103-104031}
M. Guo and S. Gao,
Universal properties of light rings for stationary axisymmetric spacetimes,
\href{https://doi.org/10.1103/PhysRevD.103.104031}
{\PRD \textbf{103}, 104031 (2021)}.

\bibitem{PRD105-024049}
M. Guo, Z. Zhong, J. Wang, and S. Gao,
Light rings and long-lived modes in quasiblack hole spacetimes,
\href{https://doi.org/10.1103/PhysRevD.105.024049}
{\PRD \textbf{105}, 024049 (2022)}.

\bibitem{PRD108-104041}
S.-P. Wu and S.-W. Wei,
Topology of light rings for extremal and nonextremal Kerr-Newman-Taub-NUT black holes without Z$_2$ symmetry,
\href{https://doi.org/10.1103/PhysRevD.108.104041}
{\PRD \textbf{108}, 104041 (2023)}.

\bibitem{2401.05495}
P.V.P. Cunha, C.A.R. Herdeiro, and J.P.A. Novo,
Light rings on stationary axisymmetric spacetimes: blind to the topology and able to coexist,
\href{https://doi.org/10.1103/PhysRevD.109.064050}
{\PRD \textbf{109}, 064050 (2024)}.

\bibitem{PRD107-064006}
S.-W. Wei and Y.-X. Liu,
Topology of equatorial timelike circular orbits around stationary black holes,
\href{https://doi.org/10.1103/PhysRevD.107.064006}
{\PRD \textbf{107}, 064006 (2023)}.

\bibitem{JCAP0723049}
X. Ye and S.-W. Wei,
Topological study of equatorial timelike circular orbit for spherically symmetric (hairy) black holes,
\href{https://doi.org/10.1088/1475-7516/2023/07/049}
{\JCAP \textbf{07} (2023) 049}.

\bibitem{2406.13270}
X. Ye and S.-W. Wei,
Novel topological phenomena of timelike circular orbits for charged test particles,
\href{https://arxiv.org/abs/2406.13270}{arXiv:2406.13270}.


\end{thebibliography}
\end{document}